\documentclass{main}

\usepackage{color}
\usepackage{lineno,hyperref}
\usepackage{amssymb}
\usepackage{amsmath}
\usepackage{xspace}

\def\beq{\begin{equation}}
\def\eeq{\end{equation}}
\def\be{\begin{eqnarray}}
\def\ee{\end{eqnarray}}

\newcommand{\Jpsi}{\ensuremath{{\rm J}/\psi}\xspace}
\newcommand{\de}[1]{{\rm d}#1}

\newcommand{\Photo}{}

\begin{document}

\title{\Large Vector meson production using the Balitsky-Kovchegov equation including the dipole orientation}

\author{J. Cepila, J. G. Contreras, M. Vaculciak}

\address{Faculty of Nuclear Sciences and Physical Engineering, Czech Technical University in Prague.\\ Brehova 7, Prague, 112519 Czech Republic}

\maketitle\abstracts{
In this proceedings a solution of the target-rapidity Balitsky-Kovchegov (BK) equation is presented considering the complete impact-parameter dependence, including the orientation of the dipole with respect to the impact-parameter vector. The target-rapidity formulation of the BK equation introduces non-locality in rapidity. Three different prescriptions are considered to take into account the rapidities preceding the initial condition. The solutions are used to compute the structure functions of the proton and the diffractive photo- and electro-production of $J/\psi$ off protons. The predictions agree well with HERA data, confirming that the target-rapidity Balitsky-Kovchegov equation with the full impact-parameter dependence is a viable tool to study the small Bjorken-$x$ limit of perturbative QCD at current facilities like RHIC and LHC as well as in future colliders like the EIC.}


\section{Introduction\label{sec:intro}}
The high-energy limit of perturbative quantum chromodynamics (pQCD) has been intensively studied in the past years. It was possible due to precise measurements from HERA experiments~\cite{Newman_2014} and from LHC experiments~\cite{Akiba_2016}. This limit is equivalent to the low Bjorken-$ x $ behavior of the gluon density in the target. Experimental data on the structure functions from DIS process~\cite{Abramowicz:2015mha} suggest that the gluon distribution rises rapidly for gluons carrying a small Bjorken-$x$. The growth of the gluon distribution is driven by splitting processes which increase the number of gluons in the proton. This mechanism was described successfully by the BFKL evolution equation~\cite{kuraev1977pomeranchuk,balitsky1978pomeranchuk}. However, when the gluon occupancy becomes large enough, recombination processes activate~\cite{Gribov:1983ivg,Mueller:1985wy} until a dynamical balance between both processes, called gluon saturation, is reached. 

One of the tools to describe the evolution of the proton structure at high energies within pQCD including gluon saturation is the Balitsky-Kovchegov (BK) equation~\cite{Balitsky:1995ub,Kovchegov:1999yj,Kovchegov:1999ua}, which describes the evolution in rapidity of the interaction between a color dipole and a hadronic target. It can be interpreted as a dressing of the original color dipole under the evolution towards higher energies by emitting additional gluons. In the limit of large number of colors the emitted gluon can be interpreted as a new dipole which effectively splits the parent dipole into two daughter dipoles. However, the BK equation also introduces a non-linear term that accounts for the possibility that two dipoles recombine.    

The original BK equation used the projectile rapidity ($ Y $) as the evolution variable. Recently, it was proposed to use the target rapidity ($\eta=\ln(x_0/x)$) as the evolution variable~\cite{Ducloue:2019ezk} in order to improve the stability of the equation by ensuring the correct time ordering of gluon emissions. Here, $x_0$ is the Bjorken-$x$ at which the BK evolution starts. However, this change of evolution variable also introduced non-local terms in the equation. 

The solution of the BK equation is the dipole scattering amplitude $N(\vec{r},\vec{b},\eta)$ which depends on two two-dimensional vectors defined in the transverse plane of the dipole. They can be decomposed into four scalar variables: $r=|\vec{r}|$ corresponding to the dipole size, $b=|\vec{b}|$ corresponding to the distance between the dipole and the target, $ \theta $ corresponding to the angle between $ \vec{r} $ and $ \vec{b} $ and $ \phi $ corresponding to the angle between $ \vec{b} $ and a fixed axis. Most of the solutions of the BK equation to date assumed a large and homogeneous target and, hence, only a dependence on $r$ was considered.  

Later on, the inclusion of the impact-parameter dependence in the solutions of the BK equation evolved with the projectile rapidity has been studied in~\cite{Golec-Biernat:2003naj}, where large Coulomb-like tails at large $ b $ were reported. However, it was shown that with some ad-hoc modifications to account for confinement, HERA data could be described~\cite{Berger:2010sh,Berger:2011ew,Berger:2012wx}. 

Soon thereafter, the kernel of the BK equation improved with resummation of some of the next-to-leading order diagrams has been published~\cite{Beuf:2014uia,Iancu:2015vea,Iancu:2015joa} and, afterwards, it has been demonstrated, that this version of the BK equation can be solved including the $r$ and $b$ dependence without the appearance of Coulomb tails, at least for energies relevant for currently available data~\cite{Cepila:2018faq,Bendova:2019psy,Cepila:2020xol,Bendova:2020hbb}. 

The suppression of Coulomb tails, which allowed for phenomenological applications of the impact-parameter dependent BK equation, was mainly due to two factors: the collinearly improved kernel and choosing an appropriate initial condition. However, the kernel of the BK equation formulated in target rapidity is missing the part that supressed the Coulomb tails when evolving with the projectile rapidity. This raises the question if the new version of the BK equation is usable for phenomenological applications.

Another natural step in the study of the solution of the BK equation is to include a non-trivial dependence on $ \theta $ and, hence, allow for inhomogeneous targets. This effect has been studied partially in different frameworks, e.g. Refs.~\cite{Hatta:2017cte,Iancu:2017fzn,Mantysaari:2020lhf,Dumitru:2021tvw,Kopeliovich:2021dgx}.

In these proceedings, a numerical solutions to the target-rapidity Balitsky-Kovchegov equation including the impact-parameter dependence as well as the dependence on the angle between the dipole size and the impact-parameter vectors is presented. The major observation is that the treatment of the non-local term for rapidities earlier than the point where the BK evolution starts influences the behavior of the Coulomb tails and offers a way to tame them. 

\section{Balitsky-Kovchegov equation in target rapidity\label{sec:overview}}

The BK equation in target rapidity has a form~\cite{Ducloue:2019ezk} 
\beq
\frac{{\rm d}N(\vec{r}, \vec{b}, \eta)}{{\rm d}\eta} = \int \mathrm{d} \vec{r}_1 K(r, r_1, r_2) \Big[ N(\vec{r}_1, \vec{b}_1, \eta_1) + N(\vec{r}_2, \vec{b}_2, \eta_2) - N(\vec{r}, \vec{b}, \eta)	- N(\vec{r}_1, \vec{b}_1, \eta_1) N(\vec{r}_2, \vec{b}_2,\eta_2) \Big]. \label{eq:BK}
\eeq
The first three terms with $ N $ on the right-hand-side of the equation take into account the splitting of a dipole at $(\vec{r},\vec{b})$ into two dipoles at $(\vec{r}_1,\vec{b}_1)$ and $(\vec{r}_2,\vec{b}_2)$, while  the last term represents the recombination of two  dipoles. 
\begin{figure}[h!]
	\centering 
	\includegraphics[width=0.3\textwidth]{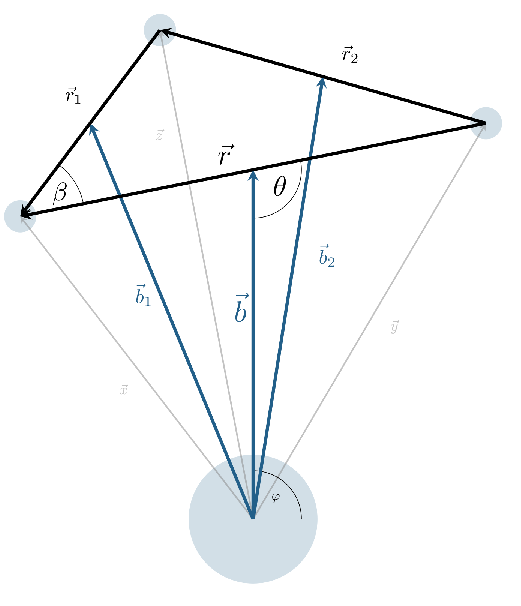}
	\caption{Kinematics of the parent and daughter dipoles during the BK equation evolution considering non-trivial dependence of the solution on the angle $ \theta $.
	\label{fig:3Dkine} }
\end{figure}
The vectors introduced above are linked through the kinematic formulas
\begin{equation}
\vec r_2=\vec r - \vec r_1\qquad \vec b_1=\vec b_{q1}-\frac{\vec r_1}{2}\qquad \vec b_{q1}=\vec b+\frac{\vec r}{2}\qquad \vec b_2=\vec b_{q2}-\frac{\vec r_2}{2}\qquad \vec b_{q2}=\vec b-\frac{\vec r}{2}.
\end{equation}
The rapidities $\eta_1$ and $\eta_2$ introduce the non-local variables defined as 
\begin{equation}
\eta_j = \eta - \max\lbrace 0, \ln(r^2/r^2_j)\rbrace.
\label{eq:shift}
\end{equation}
The collinearly improved kernel in the target rapidity is given by 
\beq
K(\Vec{r}, \Vec{r}_1, \Vec{r}_2) = \frac{\bar{\alpha}_s}{2\pi} \frac{r^2}{r_1^2r_2^2} \bigg[ \frac{r^2}{\min \lbrace r_1^2, r_2^2 \rbrace} \bigg]^{\pm \bar{\alpha}_s A_1} \label{eq:kernel},
\eeq
where the constant $A_1 = \frac{11}{12}$ and $\bar{\alpha}_s = \frac{N_{\rm C}}{\pi} \alpha_s$ with the number of colors $N_{\rm C} = 3$ and \mbox{$\alpha_s = \alpha_s(\min\lbrace r, r_1, r_2 \rbrace)$} being the running strong coupling constant 
\begin{equation}\label{alph}
	\alpha_{s, n_{f}} (r^{2}) = \frac{4\pi}{\beta_{n_{f}}\ln\left(\frac{4C^{2}}{r^{2}\Lambda ^{2}_{n_{f}}}\right)},
\end{equation}
where $n_{f}$ corresponds to the number of flavors, $ \beta_{n_f}=(11N_{\rm C}-2n_f)/3 $, $C^{2}$ is an infrared regulator adjusted to describe data and $ \Lambda ^{2}_{n_{f}} $ is a evaluated in the variable-number-of-flavors scheme~\cite{Albacete:2010sy}. In this work, the coupling is frozen at  $\alpha_s^{\rm sat}$ = 1 as in~\cite{Iancu:2015vea}.

\noindent The initial condition for the start of the evolution is given by
\begin{equation}
N(\vec{r}, \vec{b}, \eta = 0)  = 1 - \exp{\left(- \frac{1}{4}(Q_{s0}^2\,r^2)^\gamma \,T(b,r)\left\{1 + c\cos(2\theta)\right\}\right)},
\label{eq:N0}
\end{equation}
with 
\begin{equation}
T(r, b)= \exp{\left(-\frac{b^2+(r/2)^2}{2B}\right)}.
\label{eq:Trb}
\end{equation}
The parameter $Q_{s0}^2$ is related to the onset of the saturation, $T(r, b)$ corresponds to the transverse profile of the target, the parameter $ B $ is related to the size of the target and $\gamma$ is the so-called anomalous dimension (see Ref.~\cite{Albacete:2004gw}). The parameter $c$ controls the amount of the expected asymmetry on the $\theta$ dependence. 

Three different variants how to deal with shifts $ \eta_1,\eta_2 $ being negative, see Eq.~(\ref{eq:shift}), were considered:
\begin{itemize}
\item[A:] No extrapolation bellow $\eta = 0$, namely, $N(\vec{r}, \vec{b}, \eta<0) = 0$. 
\item[B:] Smooth suppression in the range $\eta=\ln(x_0/1)$ and $\eta=0$ according to the GBW model~\cite{Golec-Biernat:1998zce}
\begin{equation}
	N(\vec{r}, \vec{b}, \eta<0)  = 1 - \exp{\left(- \frac{1}{4}\left[(x_0/x)^{\lambda}\,Q_{s0}^2\,r^2\right]^\gamma \,T(b,r)\left\{1 + c\cos(2\theta)\right\}\right)},
	\label{eq:N0min}
\end{equation}
and then $N(\vec{r}, \vec{b}, \eta<\ln(x_0/1)) = 0$.  
\item[C:] Flat extrapolation to initial conditions: $N(\vec{r}, \vec{b}, \eta<0) = N(\vec{r}, \vec{b}, 0) $. 
\end{itemize}
The BK equation is solved numerically in a logarithmic grid in $r$ and $b$ and a linear grid in $\theta$ using the Runge--Kutta method with the integrals performed with Simpson's method. The step in rapidity is 0.1. The parameter values that we have used for the solution of the BK equation are $x_0=0.01$, $Q_{s0}^2=0.496$~GeV$^2$, $B=3.8$~GeV$^{-2}$, $\gamma=1.25$, $\lambda=0.288$, $c=1$ and $C=30$. The masses of the light quarks were taken to be 0.1~GeV$/c^2$ and the mass of the charm quark was taken to be 1.3~GeV$/c^2$. 

\section{Results\label{sec:results}}
\begin{figure}[!h]
\centering 
\includegraphics[width=0.39\textwidth]{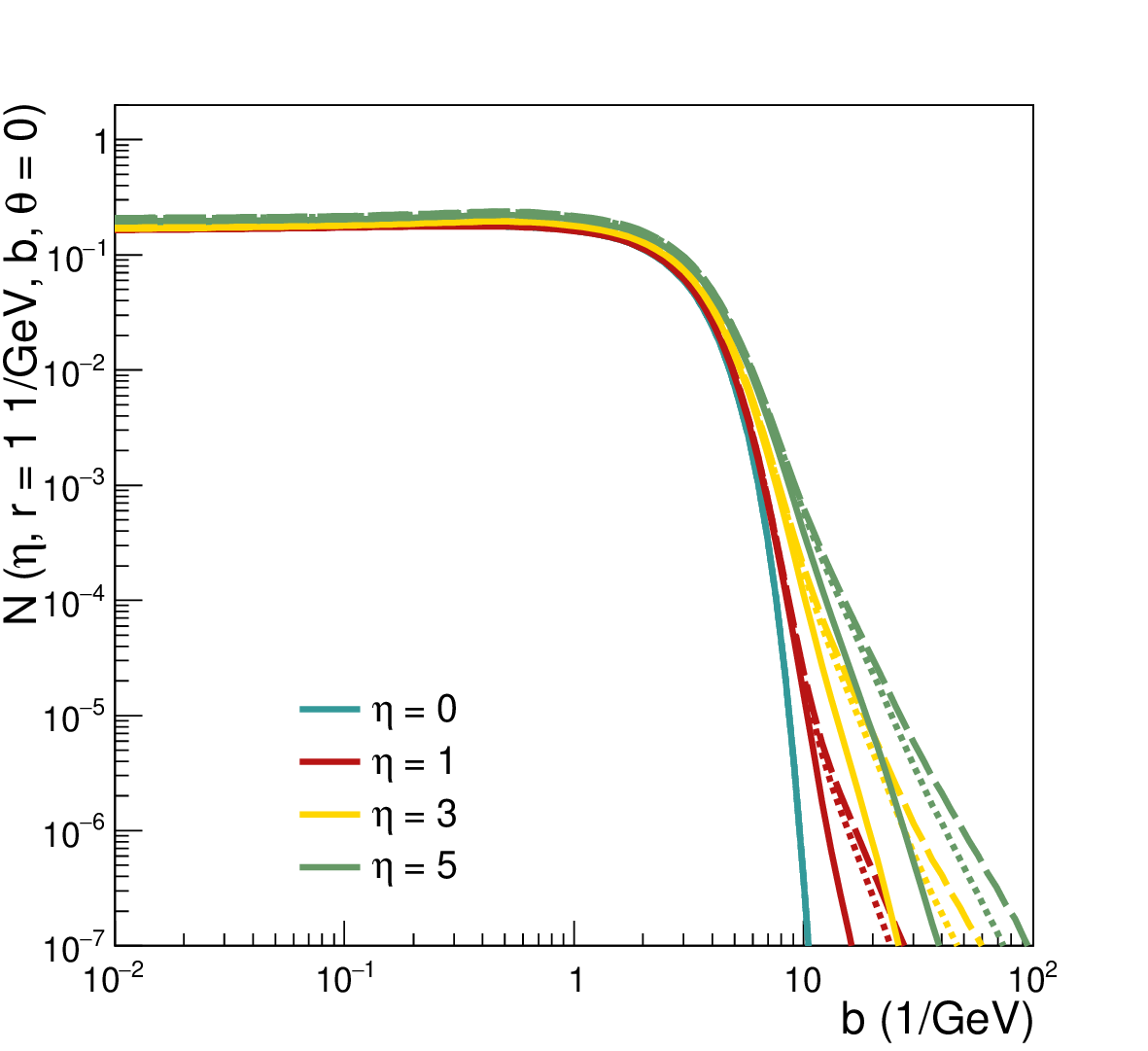}
\includegraphics[width=0.39\textwidth]{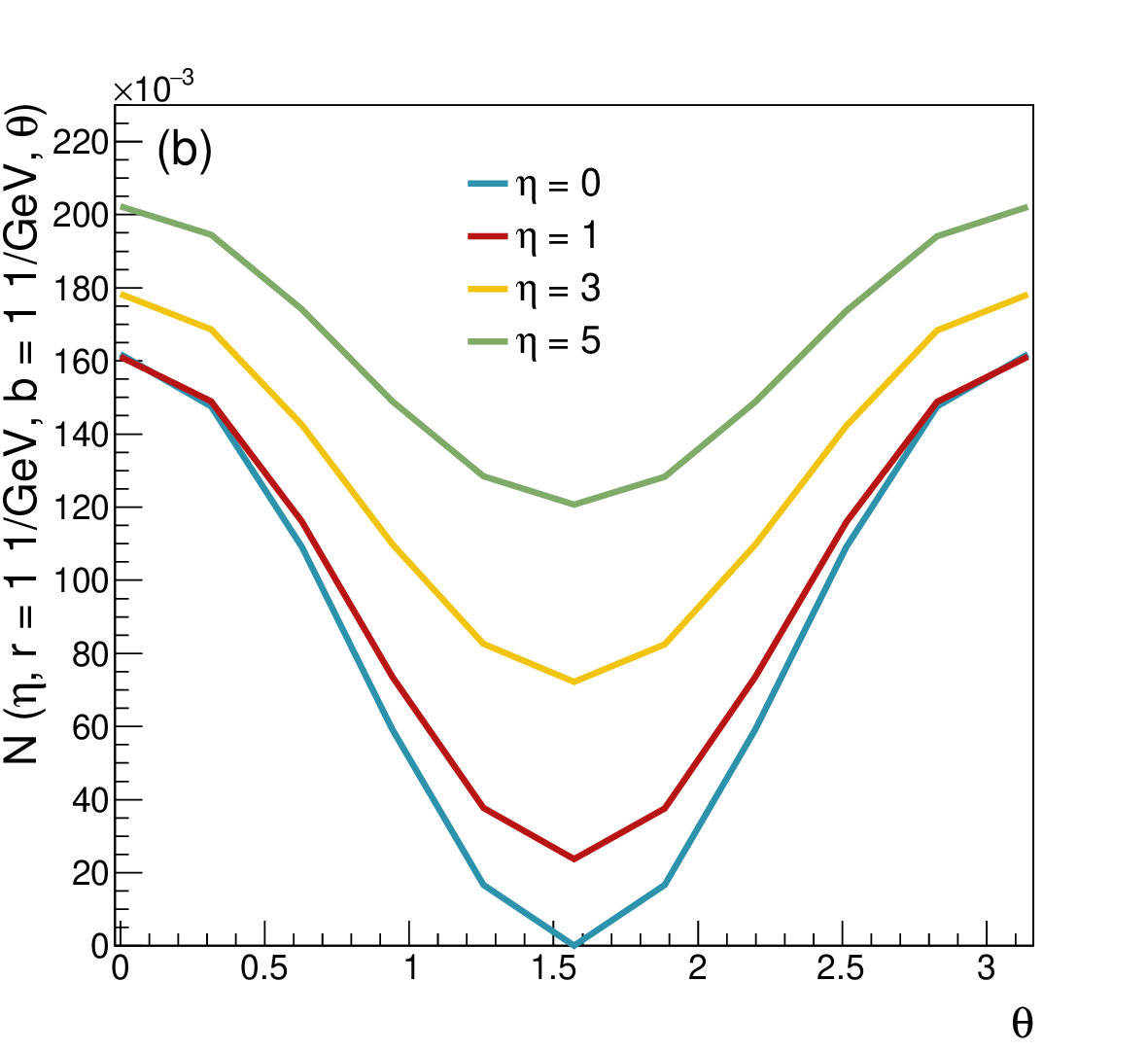}
\caption{Left: Dependence of the dipole amplitude on $b$ for fixed dipole of size $r=1$ GeV$^{-1}$ and fixed angle $\theta=0$ for three different approaches to non-locality. Solid line corresponds to the approach A, dotted line to the approach B and dashed line to the approach C. Right: Dependence of the dipole amplitude on $\theta$ for a dipole of size $r=1$~GeV$^{-1}$ at impact parameter $b=1$~GeV$^{-1}$. Solutions are shown at different rapidities for the approach A to non-locality.}\label{fig:nonlocal}
\end{figure}

The solution of the collinearly improved BK equation in target rapidity dependent explicitly on the three kinematic variables $ r,b,\theta $ has been successfully found~\cite{Cepila:2023pvh}. It shows that only one of the approaches to non-locality strongly suppresses Coulomb tails, see the left plot of Fig. \ref{fig:nonlocal}. It means that only the solution with option A is suitable for phenomenological predictions.

The dependence on $\theta$, the angle between the dipole-size and the impact-parameter  vectors, of the solutions of the BK equation is shown in the right plot of Fig.~\ref{fig:nonlocal} for different rapidities. The range of rapidities  roughly corresponds to the region that can be covered by existing experimental results or by those expected in the near and medium term. The figure presents solutions obtained with approach A to non-locality.

\begin{figure}[!h]
\centering 
\includegraphics[width=0.39\textwidth]{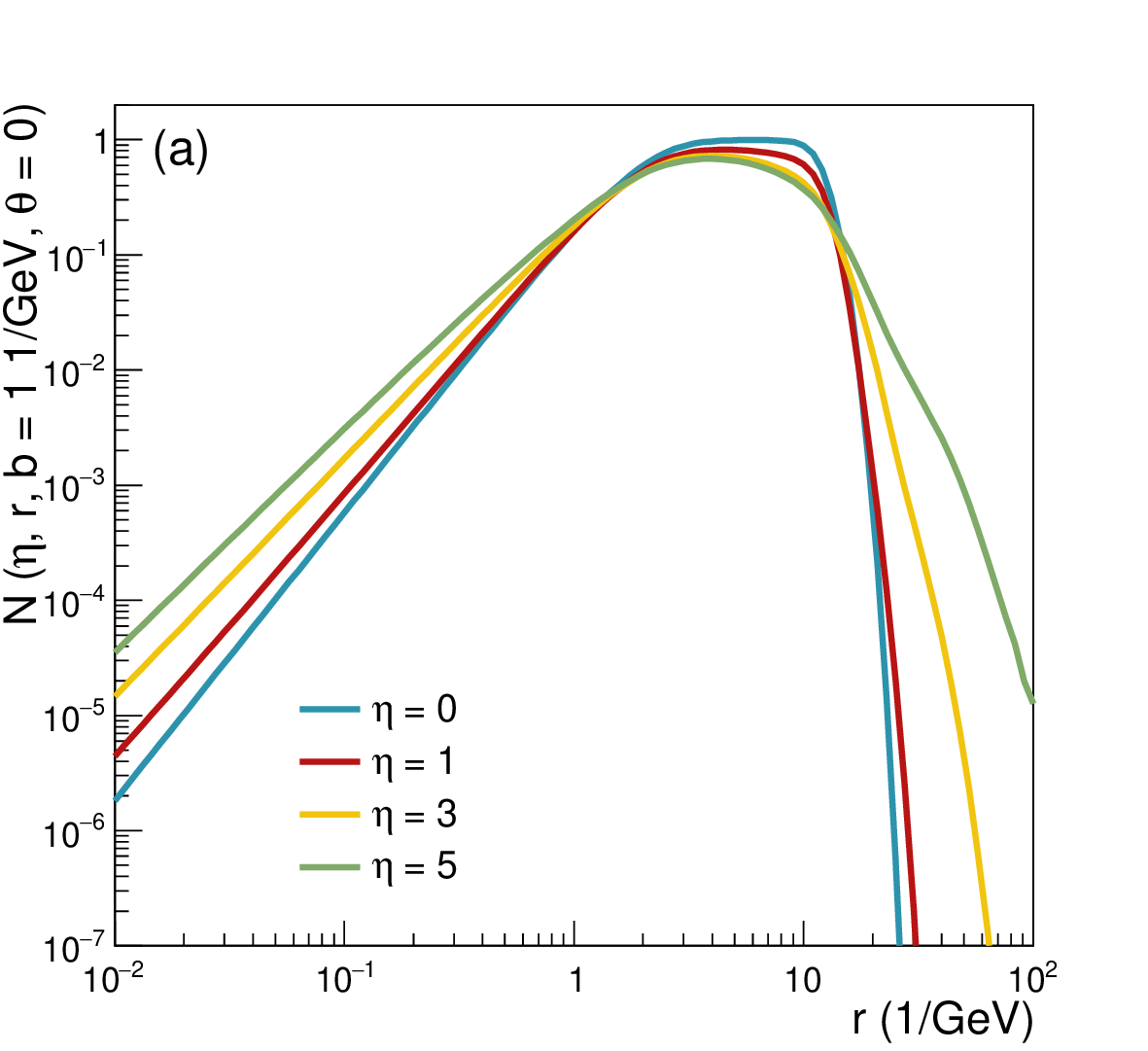}
\includegraphics[width=0.39\textwidth]{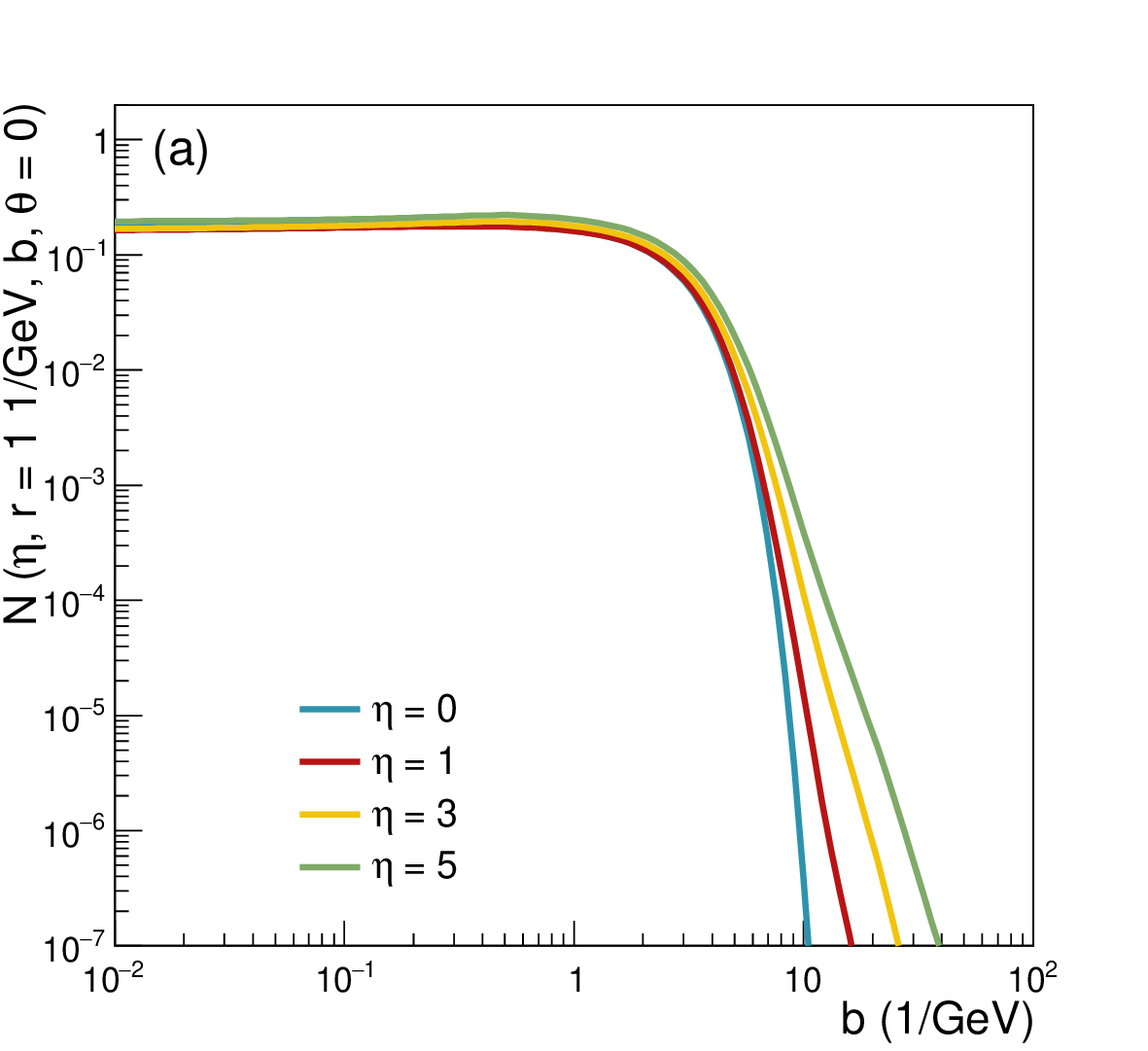}
\caption{Dependence of the dipole amplitude (left) on dipole size $r$ at an impact parameter $b=1$ GeV$^{-1}$ and (right) on impact parameter $b$ for a dipole size $r=1$  GeV$^{-1}$. The angle between the dipole-size and the impact-parameter vectors is $\theta=0$. Solutions are shown at different rapidities for the approach A.}\label{fig:rbA}
\end{figure}

The $r$ dependence of solutions of the BK equations are shown in the left plot of Fig.~\ref{fig:rbA}. The dipole amplitudes are shown at different rapidities at one impact parameter for approach A. At large values of $r$ a wave front develops, in addition to the traditional wave front towards small values of $r$. However, the large-$ r $ region is usually suppressed by the wave function of particular process and so the development of tails at large $ r $ does not spoil the usability of the solution for phenomenology. Moreover, the emergence of the tails is a direct consequence of having a target that is finite in the impact-parameter plane~\cite{Cepila:2023pvh}. 
The dependence on impact parameter $b$ of the solutions of the BK equation at fixed $\theta =0$ is shown in the right plot of Fig.~\ref{fig:rbA} at different rapidities for the approach A. The figure shows a flat behavior at small and medium impact parameters, while the dipole amplitude starts to decrease rapidly at large impact parameters because the initial condition represents a finite target. The evolution increases the range in impact parameter where the dipole amplitude is sizable, but it also changes the shape of the amplitude at large impact parameters. However, the speed with which the dipole amplitude rises towards large $ b $ is not so steep that it would result in unphysical predictions of e.g. structure functions of DIS towards low Bjorken-$ x $~\cite{Cepila:2023pvh}.

The new observation brought up by the non-locality present in the BK equation evolved in target rapidity is that the shape and size of the large-$r$ and of the large-$b$ wavefront depend on the treatment of the region of rapidities earlier than the initial rapidity of the evolution.

\section{Inclusive and exclusive observables\label{sec:app}}
The use of the solution to the collinearly improved BK equation in target rapidity is shown using the inclusive DIS process and exclusive vector meson production. The structure function $F_2$ calculated using the dipole amplitude $N$ can be written as~\cite{Cepila:2023pvh}
\begin{equation}
F_2 (x, Q^2) = \frac{Q^2}{4\pi^2 \alpha_{\rm em}}\sum\limits_{f} \left[ \sigma^{\gamma*p}_{L, f}\left(Q^2, x_f(x, Q^2)\right) + \sigma^{\gamma*p}_{T, f}\left(Q^2, x_f(x, Q^2)\right) \right],
\end{equation}
where $f$ denotes the flavour of a quark, $Q^2$ is the virtuality of the exchanged photon, $ \alpha_{\rm em}$ is the electromagnetic coupling constant, 
\begin{equation}
x_f = \frac{x_0 e^{-\eta}}{1+4\frac{m_{f}^2}{Q^2}}
\end{equation}
and 
\begin{equation}
\sigma^{\gamma*p}_{L, T, f}(Q^2, x_f) = 4\pi\int\de{r} r \int\de{z} |\psi_{L, T, f}(r, z, Q^2)|^2 \int \de{b} b\int\limits_{0}^{\pi} \de{\theta} 2N(r, b, \theta, \eta(x_f)),
\end{equation}
with the longitudinal ($L$) and transverse ($T$) light-cone wave functions
\begin{equation}
\left| \psi_{L, f}(r, z, Q^2) \right|^2= \frac{N_C \alpha_{em}}{2\pi^2}e_f^2 4 Q^2 z^2 (1-z)^2 K_0^2\left(r \epsilon\right) 
\end{equation}
and
\begin{align}    
\left|\psi_{T, f}(r, z, Q^2)\right|^2 = \frac{N_C \alpha_{em}}{2\pi^2}e_f^2 \left[(z^2 + (1-z)^2) \epsilon^{2} K_1^2\left(r \epsilon\right)+ m_{f}^2 K_0^2\left(r \epsilon\right)
\right]. 
\end{align}
where $\epsilon=\sqrt{z(1-z)Q^2 + m_{f}^2}$, and $K_{0,1}$ are Bessel functions.

Diffractive exclusive vector meson production calculated using the dipole amplitude $N$ is given by the sum of the transverse and longitudinal contributions:
\begin{equation}
\frac{\de \sigma_{T,L}}{\de |t|}(t, Q^2, W) = \frac{1}{16\pi} (1 + \beta_{T,L}^2) R^2_{L,T} \left| \mathcal{A} _{T,L} \right|^2, 
\end{equation}
where 
\begin{equation}
\mathcal{A}(t, Q^2, W) = i 2\pi\int \de r r \int\limits_{0}^{1} \frac{\de z}{4\pi}  \left( \Psi_V^\dagger \Psi\right)_{T,L}\int \de^2 \vec{b}\,e^{-i[\vec{b}- (\frac{1}{2}-z)\vec{r}]\vec{\Delta}} 2 N\left(r,b,\theta, \eta)\right)
\end{equation}
with $x=(Q^2+M^2)/(W^2+Q^2)$, $M$ the mass of the vector meson, $W$ the center-of-mass energy of the photon--proton system, $\vec{\Delta}^2=-t$, and the wave function of the vector mesons given by the boosted Gaussian model with the values of all parameters as in Ref.~\cite{Bendova:2018bbb}. The corrections to the real part of the amplitude and to the skewedness effect are 
\begin{equation}
\beta_{T,L} = \tan (\pi \lambda_{T,L}/2), \qquad R_{T,L} = \frac{2^{2\lambda_{T,L}+3}}{\sqrt{\pi}} \frac{\Gamma (\lambda_{T,L} + 5/2)}{\Gamma (\lambda_{T,L} + 4)},\qquad \lambda_{T,L} = \frac{\partial \ln \mathcal{A}_{T,L}(t=0)}{\partial \ln (1/x)}.
\end{equation}

\section{Predictions of observables\label{sec:pheno}}
We have compared in \cite{Cepila:2023pvh} the solutions to the collinearly improved BK equation evolved in target rapidity to HERA measurements of structure functions obtained in deep-inelastic scattering~\cite{H1:2009pze} and the $ t $-distribution of the diffractive exclusive $\Jpsi$ vector meson photo- and electro-production~\cite{H1:2005dtp}. 
\begin{figure}[!h]
\centering 
\includegraphics[width=0.39\textwidth]{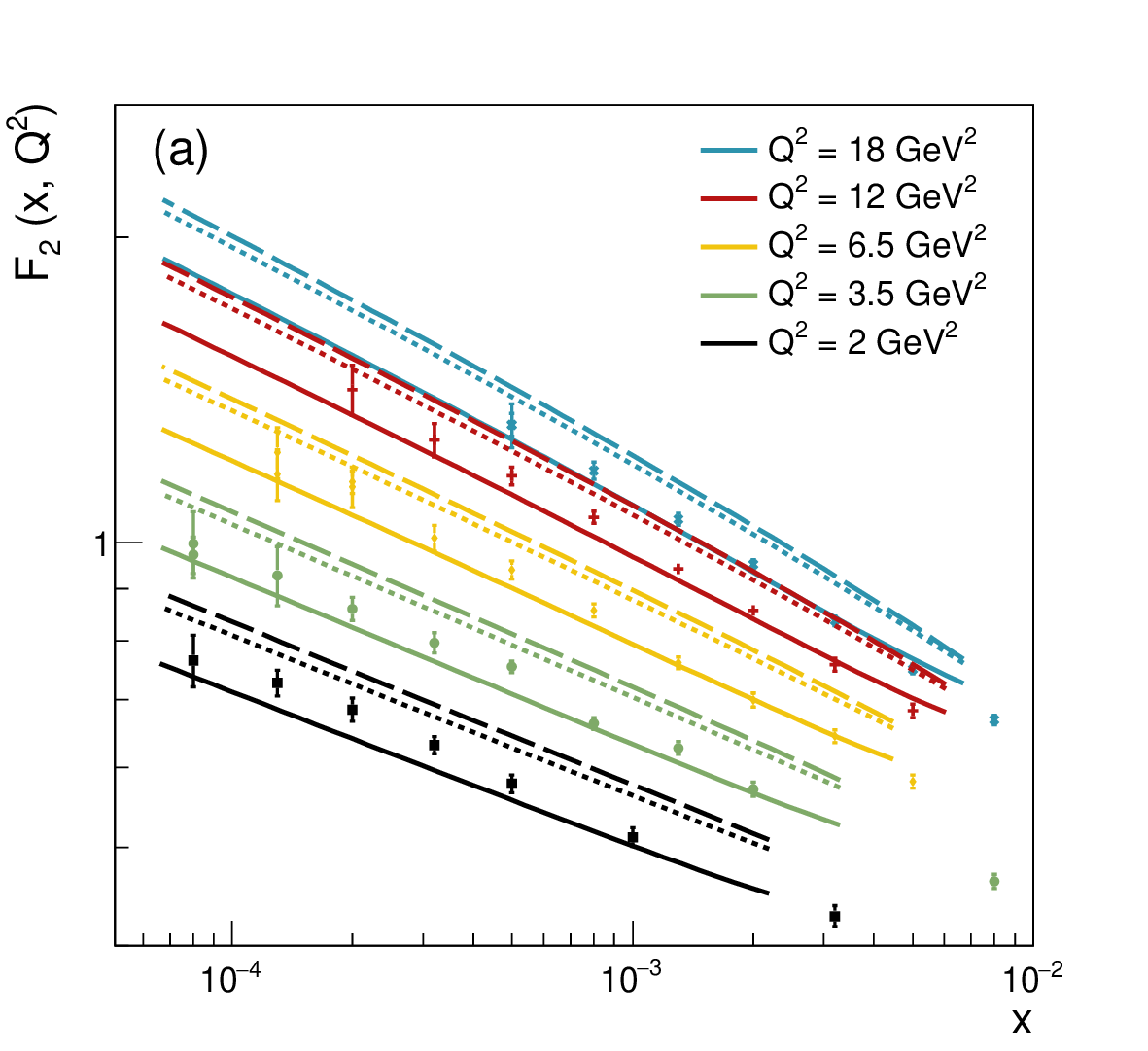}
\includegraphics[width=0.39\textwidth]{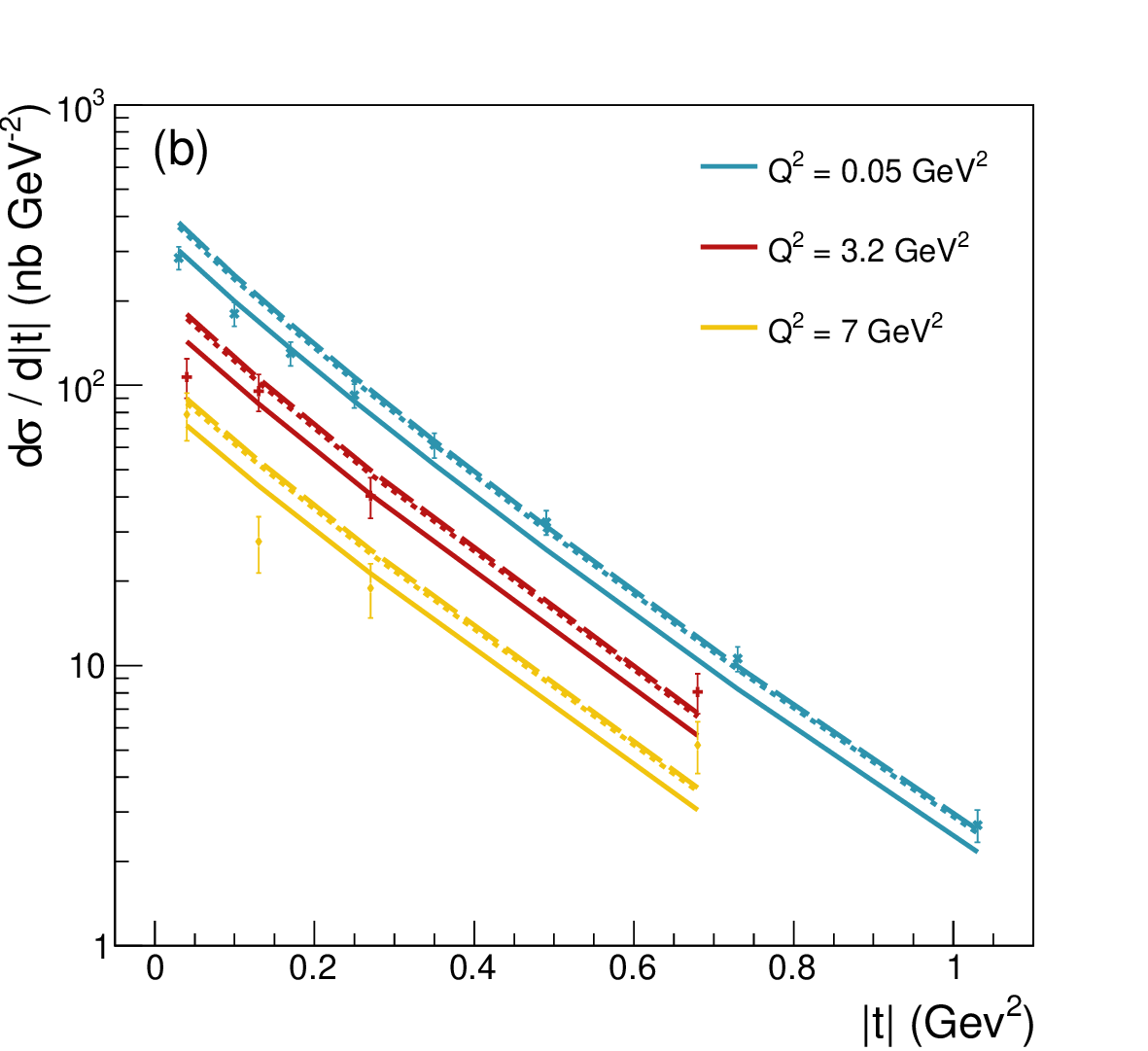}
\caption{Predictions of inclusive and exclusive observables using the target rapidity BK equations solved with the inclusion of angular correlations between the dipole orientation and the impact parameter. The three approaches to deal with the non-localities for early rapidities are shown with solid (A), dotted (B) and dashed (C) lines. Left plot shows the $F_2$ structure function as measured at HERA and right plot shows the cross section for diffractive exclusive $\Jpsi$ vector meson photo- and electro-production.}\label{fig:pheno}
\end{figure}
The comparison of the predictions of the three approaches to deal with the non-locality of the BK equation in target rapidity for early rapidities with data from HERA is shown in Fig.~\ref{fig:pheno}. In all cases shown in the figure, approaches B and C predict a larger cross section than approach A. The $F_2$ data is reasonably well described by approach A. The emergence of Coulomb tails for approaches B and C is not that fast to spoil the usability of the BK equation solution for the description of current data for $F_2$. However, at lower Bjorken-$ x $ the use of the approach B and C would be more and more problematic. The same can be said about the comparison to the vector meson data. This demonstrates that the BK equation in target rapidity with the A approach can be used for phenomenological applications without adding ad-hoc prescriptions to deal with the influence of Coulomb tails. 

\section{Summary and outlook\label{sec:summary}}
In these proceedings a new solutions of the BK equation evolved in the target rapidity including the dependence on the size of the dipole $ r $, the magnitude of the impact-parameter $ b $ and the angle between those two vectors $ \theta $ is presented. A new form of initial conditions have been proposed that are inspired by the GBW model in $ r $-dependence and respects the $ b $- and $ \theta $-dependence from recent models. The kernel of the target rapidity BK equation lacks the Coulomb-tails suppressing term. However, we have shown that the presence of non-local terms in the evolution together with the particular choice of the approach to the contributions from terms evaluated at rapidities before the evolution starts effectively suppresses the Coulomb tails. Also, the solutions have been used to obtain predictions for physical observables for inclusive process, namely the $F_2$ structure function of protons, and for exclusive process, namely the cross section for diffractive exclusive photo- and electro-production of $\Jpsi$ vector mesons off protons.  Both sets of predictions are compared to existing data from HERA and a reasonable agreement is found without any ad-hoc corrections. 
This opens the possibility to use solutions of this equation to explore other observables that are to be measured at current facilities, like RHIC and the LHC, or those that will enter operation in the near future, like the EIC.

\section*{Acknowledgments}

This work was partially funded by the Czech Science Foundation (GAČR), project No. 22-27262S.

\section*{References}
\bibliography{bibliography}

\begin{thebibliography}{10}

\bibitem{Newman_2014}
Paul~R. Newman and Matthew Wing.
\newblock The hadronic final state at hera.
\newblock {\em Reviews of Modern Physics}, 86(3):1037–1092, August 2014.

\bibitem{Akiba_2016}
K~Akiba et~al.
\newblock Lhc forward physics.
\newblock {\em Journal of Physics G: Nuclear and Particle Physics},
  43(11):110201, October 2016.

\bibitem{Abramowicz:2015mha}
H.~Abramowicz et~al.
\newblock {Combination of measurements of inclusive deep inelastic ${e^{\pm
  }p}$ scattering cross sections and QCD analysis of HERA data}.
\newblock {\em Eur. Phys. J.}, C75(12):580, 2015.

\bibitem{kuraev1977pomeranchuk}
Eh~A Kuraev, LN~Lipatov, and Victor~S Fadin.
\newblock Pomeranchuk singularity in nonabelian gauge theories.
\newblock {\em Sov. Phys.-JETP (Engl. Transl.);(United States)}, 45(2), 1977.

\bibitem{balitsky1978pomeranchuk}
II~Balitsky and LN~Lipatov.
\newblock The pomeranchuk singularity, process, theory.
\newblock {\em Soviet Journal of Nuclear Physics}, 28:822--822, 1978.

\bibitem{Gribov:1983ivg}
L.~V. Gribov, E.~M. Levin, and M.~G. Ryskin.
\newblock {Semihard Processes in QCD}.
\newblock {\em Phys. Rept.}, 100:1--150, 1983.

\bibitem{Mueller:1985wy}
Alfred~H. Mueller and Jian-wei Qiu.
\newblock {Gluon Recombination and Shadowing at Small Values of x}.
\newblock {\em Nucl. Phys. B}, 268:427--452, 1986.

\bibitem{Balitsky:1995ub}
I.~Balitsky.
\newblock {Operator expansion for high-energy scattering}.
\newblock {\em Nucl. Phys. B}, 463:99--160, 1996.

\bibitem{Kovchegov:1999yj}
Yuri~V. Kovchegov.
\newblock {Small x F(2) structure function of a nucleus including multiple
  pomeron exchanges}.
\newblock {\em Phys. Rev. D}, 60:034008, 1999.

\bibitem{Kovchegov:1999ua}
Yuri~V. Kovchegov.
\newblock {Unitarization of the BFKL pomeron on a nucleus}.
\newblock {\em Phys. Rev. D}, 61:074018, 2000.

\bibitem{Ducloue:2019ezk}
B.~Duclou\'e, E.~Iancu, A.~H. Mueller, G.~Soyez, and D.~N. Triantafyllopoulos.
\newblock {Non-linear evolution in QCD at high-energy beyond leading order}.
\newblock {\em JHEP}, 04:081, 2019.

\bibitem{Golec-Biernat:2003naj}
Krzysztof~J. Golec-Biernat and A.~M. Stasto.
\newblock {On solutions of the Balitsky-Kovchegov equation with impact
  parameter}.
\newblock {\em Nucl. Phys. B}, 668:345--363, 2003.

\bibitem{Berger:2010sh}
Jeffrey Berger and Anna Stasto.
\newblock {Numerical solution of the nonlinear evolution equation at small x
  with impact parameter and beyond the LL approximation}.
\newblock {\em Phys. Rev. D}, 83:034015, 2011.

\bibitem{Berger:2011ew}
Jeffrey Berger and Anna~M. Stasto.
\newblock {Small x nonlinear evolution with impact parameter and the structure
  function data}.
\newblock {\em Phys. Rev. D}, 84:094022, 2011.

\bibitem{Berger:2012wx}
Jeffrey Berger and Anna~M. Stasto.
\newblock {Exclusive vector meson production and small-x evolution}.
\newblock {\em JHEP}, 1301:001, 2013.

\bibitem{Beuf:2014uia}
Guillaume Beuf.
\newblock {Improving the kinematics for low-$x$ QCD evolution equations in
  coordinate space}.
\newblock {\em Phys. Rev. D}, 89(7):074039, 2014.

\bibitem{Iancu:2015vea}
E.~Iancu, J.~D. Madrigal, A.~H. Mueller, G.~Soyez, and D.~N.
  Triantafyllopoulos.
\newblock {Resumming double logarithms in the QCD evolution of color dipoles}.
\newblock {\em Phys. Lett. B}, 744:293--302, 2015.

\bibitem{Iancu:2015joa}
E.~Iancu, J.~D. Madrigal, A.~H. Mueller, G.~Soyez, and D.~N.
  Triantafyllopoulos.
\newblock {Collinearly-improved BK evolution meets the HERA data}.
\newblock {\em Phys. Lett. B}, 750:643--652, 2015.

\bibitem{Cepila:2018faq}
J.~Cepila, J.~G. Contreras, and M.~Matas.
\newblock {Collinearly improved kernel suppresses Coulomb tails in the
  impact-parameter dependent Balitsky-Kovchegov evolution}.
\newblock {\em Phys. Rev. D}, 99(5):051502, 2019.

\bibitem{Bendova:2019psy}
D.~Bendova, J.~Cepila, J.~G. Contreras, and M.~Matas.
\newblock {Solution to the Balitsky-Kovchegov equation with the collinearly
  improved kernel including impact-parameter dependence}.
\newblock {\em Phys. Rev. D}, 100(5):054015, 2019.

\bibitem{Cepila:2020xol}
J.~Cepila, J.~G. Contreras, and M.~Matas.
\newblock {Predictions for nuclear structure functions from the
  impact-parameter dependent Balitsky-Kovchegov equation}.
\newblock {\em Phys. Rev. C}, 102(4):044318, 2020.

\bibitem{Bendova:2020hbb}
D.~Bendova, J.~Cepila, J.~G. Contreras, and M.~Matas.
\newblock {Photonuclear $J/\psi$ production at the LHC: Proton-based versus
  nuclear dipole scattering amplitudes}.
\newblock {\em Phys. Lett. B}, 817:136306, 2021.

\bibitem{Hatta:2017cte}
Yoshitaka Hatta, Bo-Wen Xiao, and Feng Yuan.
\newblock {Gluon Tomography from Deeply Virtual Compton Scattering at Small-x}.
\newblock {\em Phys. Rev. D}, 95(11):114026, 2017.

\bibitem{Iancu:2017fzn}
Edmond Iancu and Amir~H. Rezaeian.
\newblock {Elliptic flow from color-dipole orientation in pp and pA
  collisions}.
\newblock {\em Phys. Rev. D}, 95(9):094003, 2017.

\bibitem{Mantysaari:2020lhf}
Heikki M\"antysaari, Kaushik Roy, Farid Salazar, and Bj\"orn Schenke.
\newblock {Gluon imaging using azimuthal correlations in diffractive scattering
  at the Electron-Ion Collider}.
\newblock {\em Phys. Rev. D}, 103(9):094026, 2021.

\bibitem{Dumitru:2021tvw}
Adrian Dumitru, Heikki M\"antysaari, and Risto Paatelainen.
\newblock {Color charge correlations in the proton at NLO: Beyond geometry
  based intuition}.
\newblock {\em Phys. Lett. B}, 820:136560, 2021.

\bibitem{Kopeliovich:2021dgx}
B.~Z. Kopeliovich, M.~Krelina, and J.~Nemchik.
\newblock {Electroproduction of heavy quarkonia: significance of dipole
  orientation}.
\newblock {\em Phys. Rev. D}, 103(9):094027, 2021.

\bibitem{Albacete:2010sy}
Javier~L. Albacete, Nestor Armesto, Jose~Guilherme Milhano, Paloma
  Quiroga-Arias, and Carlos~A. Salgado.
\newblock {AAMQS: A non-linear QCD analysis of new HERA data at small-x
  including heavy quarks}.
\newblock {\em Eur. Phys. J. C}, 71:1705, 2011.

\bibitem{Albacete:2004gw}
J.~L. Albacete, N.~Armesto, J.~G. Milhano, C.~A. Salgado, and U.~A. Wiedemann.
\newblock {Numerical analysis of the Balitsky-Kovchegov equation with running
  coupling: Dependence of the saturation scale on nuclear size and rapidity}.
\newblock {\em Phys. Rev. D}, 71:014003, 2005.

\bibitem{Golec-Biernat:1998zce}
Krzysztof~J. Golec-Biernat and M.~Wusthoff.
\newblock {Saturation effects in deep inelastic scattering at low Q**2 and its
  implications on diffraction}.
\newblock {\em Phys. Rev. D}, 59:014017, 1998.

\bibitem{Cepila:2023pvh}
J.~Cepila, J.~G. Contreras, and M.~Vaculciak.
\newblock {Solutions to the Balitsky-Kovchegov equation including the dipole
  orientation}.
\newblock {\em Phys. Lett. B}, 848:138360, 2024.

\bibitem{Bendova:2018bbb}
D.~Bendova, J.~Cepila, and J.~G. Contreras.
\newblock {Dissociative production of vector mesons at electron-ion colliders}.
\newblock {\em Phys. Rev. D}, 99(3):034025, 2019.

\bibitem{H1:2009pze}
F.~D. Aaron et~al.
\newblock {Combined Measurement and QCD Analysis of the Inclusive e+- p
  Scattering Cross Sections at HERA}.
\newblock {\em JHEP}, 01:109, 2010.

\bibitem{H1:2005dtp}
A.~Aktas et~al.
\newblock {Elastic J/psi production at HERA}.
\newblock {\em Eur. Phys. J. C}, 46:585--603, 2006.

\end{thebibliography}

\end{document}